\documentclass[a4paper,11pt]{article}
\usepackage{pos}

\usepackage{graphicx}
\usepackage{caption}
\usepackage{amsmath}
\usepackage{subcaption}
\usepackage{hyperref}
\usepackage{mathtools}
\usepackage{float}
\usepackage{rotating}
\usepackage{url}
\usepackage[font=scriptsize,labelfont=bf,skip=5pt]{caption}
\usepackage[T1]{fontenc}

\title{Searches for long-lived particles with the ANUBIS experiment}

\author*[a]{Aashaq Shah}

\author[a]{\textnormal{for the ANUBIS collaboration}}

\affiliation[a]{Department of Physics, Cavendish Laboratory, University of Cambridge\\
  J J Thomson Avenue, Cambridge, CB3 0HE, United Kingdom}

\emailAdd{aashaq.shah@cern.ch}

\abstract{In recent years, there has been growing interest in the search for long-lived particles (LLPs), as predicted by various extensions of the Standard Model (SM). The AN Underground Belayed In-Shaft search experiment (ANUBIS) was proposed to search for such particles by instrumenting CERN's ATLAS underground cavern with tracking detectors. This report provides an overview of the current efforts to realize the ANUBIS project focusing on the latest optimized detector geometry and the installation of proANUBIS -- a prototype or proof-of-concept demonstrator. The latter aims to offer insights into anticipated backgrounds for the ANUBIS experiment and demonstrate the feasibility of such a project. The ongoing efforts are needed to contribute to the continuous optimization and development of the ANUBIS project.}

\FullConference{The European Physical Society Conference on High Energy Physics (EPS-HEP2023)\\
 21-25 August 2023\\
Hamburg, Germany\\}

\begin{document}
\maketitle

\section{Introduction}
\vspace{-5pt}
Long-lived particles (LLPs) represent an intriguing aspect of theoretical physics~\cite{LLP_Alimena}, offering potential solutions to fundamental mysteries such as naturalness, dark matter, neutrino masses, and the baryon asymmetry of the Universe. The lifetime of LLPs can span from the micrometre scale up to the Big Bang Nucleosynthesis limit of approximately $10^{7}$ meters, making them versatile free parameters in beyond the Standard Models (BSM) of particle physics. However, neutral LLPs with longer lifetimes present a unique challenge for detection, as the sensitivity of conventional Large Hadron Collider (LHC) main detectors is hampered by their finite volume, intricate backgrounds, trigger complexities, and limited acceptance~\cite{CMS_LLP_muonPairs, CMS_LLP_EndcapMuonDetect}.

The ANUBIS experiment~\cite{ANUBIS_proposal} was proposed to complement and extend the exciting off-axis LLP search programme at CERN. Its central idea is to reduce the civil engineering costs by repurposing the existing infrastructure of the ATLAS experiment~\cite{ATLAS_experiment} at the LHC. Different geometries that include four tracking stations in ATLAS service shafts or installing tracking stations on the ceilings of ATLAS underground cavern, or installing both service shafts as well as ceilings have been recently explored. 

\begin{figure}[H]
    \centering
    \begin{subfigure}[b]{0.85\textwidth}
        \includegraphics[width=5cm, height=6.5cm]{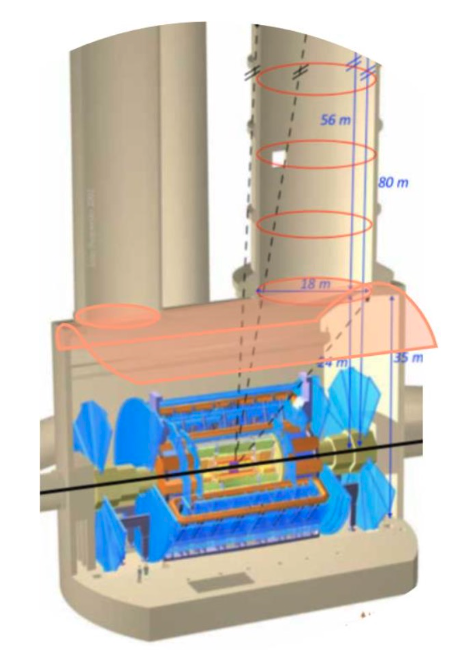} 
        \includegraphics[width=5cm, height=6.5cm]{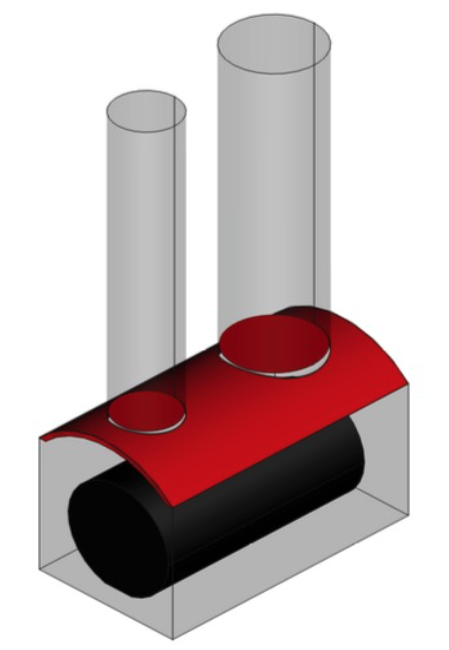} 
    \end{subfigure}
    \caption{(left) The sketch depicts the layout of the underground cavern at LHC Point 1, featuring the ATLAS experiment represented. Additionally, the PX14 and PX16 access shafts are shown. The orange colour highlighted area illustrates the current configuration of the ANUBIS detectors, to be positioned on the ceilings of the ATLAS underground cavern. Inset within the sketch are two disks covering the shafts which are going to be an integral part of the ANUBIS experiment. (right) A simplified version of the figure presented on the left side with the ANUBIS project highlighted in red and ATLAS in black colours, respectively.} \label{fig:ANUBISCeilingSketch}
\end{figure}

The simulation studies indicate that the ANUBIS project holds promising potential for the identification of novel physics when employing tracking detectors installed on the ceiling of an underground cavern as depicted in Figure~\ref{fig:ANUBISCeilingSketch}. This implementation would significantly enhance the ATLAS experiment's capability to detect the decays of neutral BSM LLPs. The project is poised to be responsive to decay scenarios situated between the vertexing limit of the ATLAS detectors and the cavern ceiling. Preliminary studies suggest that in the pursuit of these decays, ANUBIS would exhibit sensitivity to exotic phenomena significantly surpassing that of the ATLAS experiment's existing detectors. This sensitivity level matches or even exceeds that of similar proposals aimed at enhancing the sensitivity of the LHC experiments. Consequently, the ANUBIS project is anticipated to present a substantial improvement in the LHC's exploration of BSM phenomena, all achieved at a reasonable civil engineering cost.

\vspace{-5pt}
\vspace{-5pt}
The optimised proposed configuration is to instrument the ceiling of the ATLAS cavern with tracking detectors and include one disk in each of PX14 and PX16 service shafts, the area is highlighted in red colour in Figure 1. It is worth noting that the disks in the shafts containing tracking layers will be removed during the Year End Technical Stops (YETS) and other technical shutdowns to allow access to the ATLAS cavern. The motivation to instrument the ceiling, as mentioned, is because it shows more sensitivity and acceptance compared to the shaft geometry~\cite{ANUBIS_proposal}. The combination of ANUBIS's large detector volume and the adjacency to the ATLAS experiment gives rise to larger acceptance compared to other proposals~\cite{MATHUSLA_proposal} for a significant range of lifetimes, and the possibility to synchronize and fully integrate ANUBIS with ATLAS. This would result in a tracking volume that extends approximately 20 m away from the interaction point providing LLP lifetime sensitivity to $10^{-1} < c\tau~(m) < 10^{6}$. Furthermore, full integration of ANUBIS with ATLAS would allow ANUBIS to trigger the readout of ATLAS, which is technically feasible given the ATLAS Level 0 trigger latency is 5 $\mu$s at the HL-LHC~\cite{ATLAS_TDAQ_phase2}. 
\vspace{-5pt}
\section{The proANUBIS Demonstrator}
\vspace{-5pt}
To assess the feasibility of the ANUBIS project, a prototype demonstrator, proANUBIS, has been installed at Point 1 of CERN LHC~\cite{proANUBIS_setup}. This prototype is designed to validate simulation studies and contribute valuable insights into the expected backgrounds for the ANUBIS experiment. Additionally, it offers an opportunity to evaluate the detectors capabilities and functionalities in the challenging environment of the ATLAS underground cavern. 
\vspace{-10pt}
\begin{figure}[H]
    \centering
    \begin{subfigure}[b]{0.95\textwidth}
        \includegraphics[width=5.2cm, height=5.2cm]{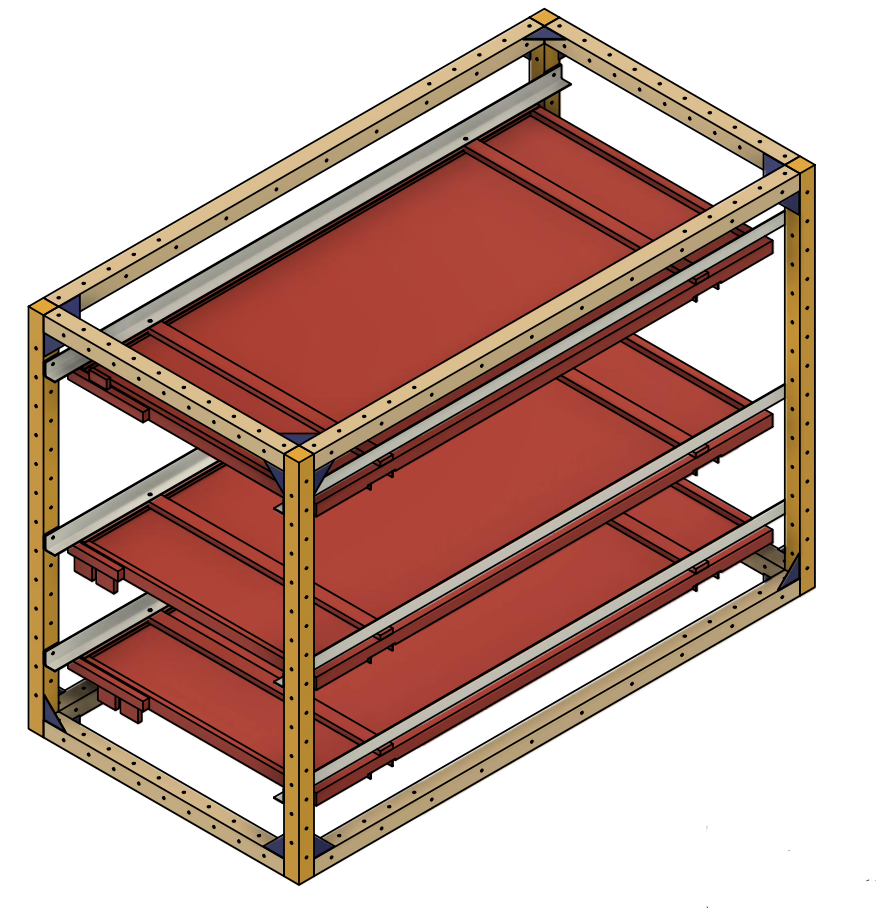} 
        \includegraphics[width=7cm, height=5.1cm]{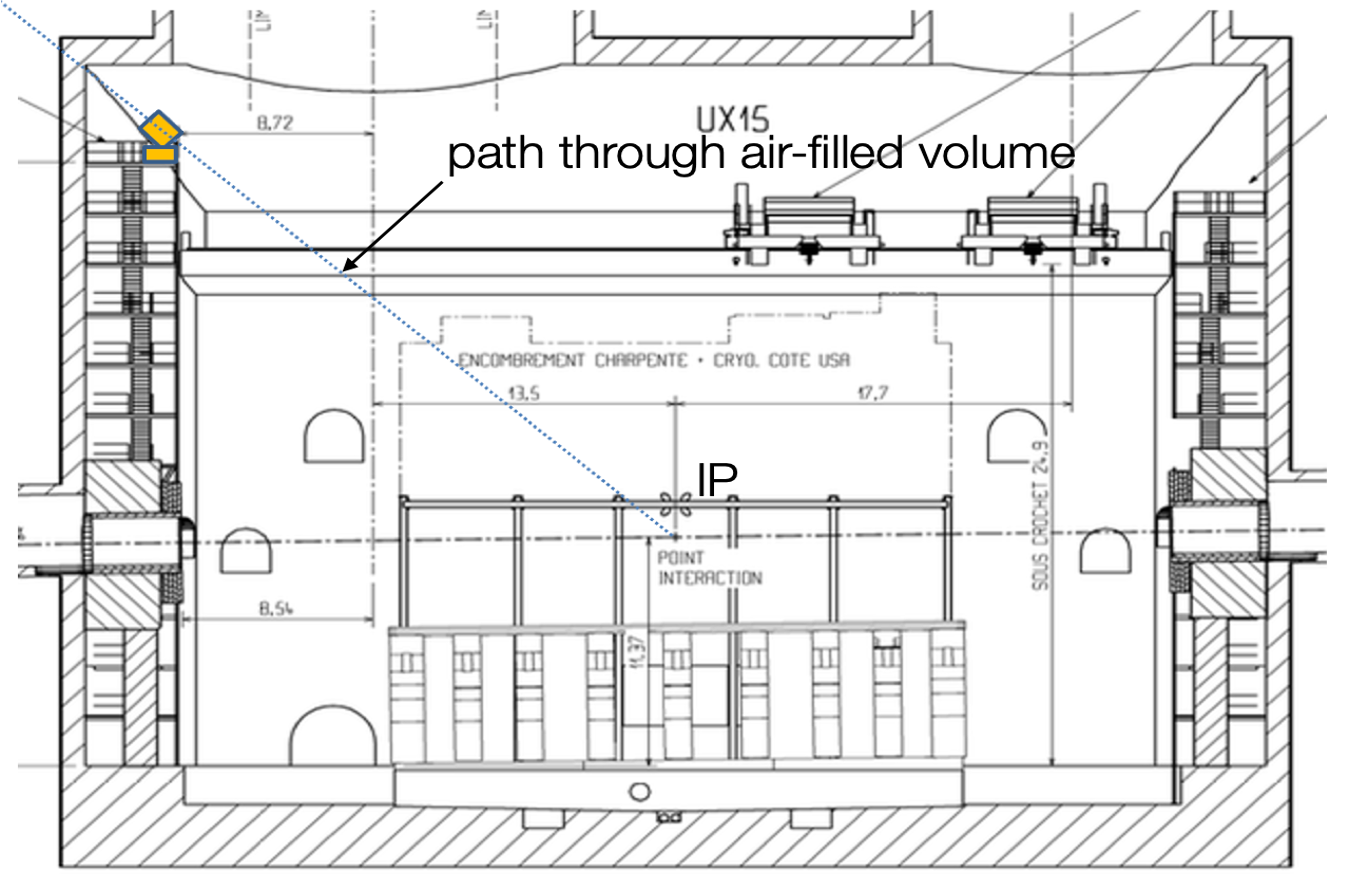} 
    \end{subfigure}
    \caption{(left) The design of the proANUBIS demonstrator is depicted, showcasing the arrangement of the three integrated tracking layers within the proANUBIS setup. At the bottom, there is a triplet configuration of RPC chambers, followed by the singlet in the middle, and the doublet on top. (right) The sketch depicting the proANUBIS setup (marked as yellow) and the angle at which it is facing the collision interaction point, the location is at level 12 side-A of the ATLAS underground cavern.} \label{fig:proANUBIS_setup}
\end{figure}
\vspace{-10pt}
The demonstrator's performance in recording background signals and potential sources of interference will provide firsthand experience. This examination aims to yield insights into mechanical engineering requirements, detector design, and readout implementations contributing to the ongoing development of the ANUBIS project.

\vspace{-5pt}
\subsection{Design and Detector Geometry}
\vspace{-5pt}
The design and geometry of the proANUBIS detector are depicted in Figure 2. The detector is composed of three tracking layers, each measuring an area of 1 m x 1.8 m and separated by a distance of approximately half a meter. The bottom layer is configured with three tracking detectors  (triplet), the top layer comprises two detectors (doublet), and the middle layer features a single detector (singlet). The decision to employ a triplet and doublet configuration is motivated by the necessity to efficiently capture muon hits resulting from LHC collisions. While these layers alone would be sufficient for muon tracking, the inclusion of the singlet layer plays a role in detecting and tracking these particles, providing an additional hit and aiding in vertexing. This configuration enhances the detector's capabilities, facilitating a more comprehensive study of potential background signals. The spacing between layers is optimized to ensure effective particle tracking within the detector volume. The overall geometry of proANUBIS is engineered to strike a balance between performance and practicality, considering the constraints imposed by the geometry and environment of the ATLAS underground cavern.
\vspace{-5pt}
\subsection{Detector Technology}
\vspace{-5pt}
The proANUBIS employs detector technology based on the latest generation of Resistive Plate Chambers, specifically the BIS78 variant~\cite{RPCUpgrade_BIS78, BIS78_electronics, ATLAS_Muon_TDR_phase2}. The primary consideration in opting for this technology is its cost-effectiveness relative to alternative detector technologies. Given ANUBIS's requirement for a larger detector area/volume, the BIS78 technology aligns perfectly with the project's aspirations.  Additionally, this technology offers good timing and spatial resolution, approximately 250 ps and a few mm, respectively, rendering it among the optimal choices for investigating the lifetime of  LLPs.
\vspace{-5pt}
\subsection{Commissioning and Installation}
\vspace{-5pt}
The commissioning and installation of proANUBIS were initiated at CERN BB5, where integration and testing procedures were conducted. The triplet, singlet, and doublet RPC chambers were inserted into the metallic frame, adhering to the design layout outlined in Figure 2. Simultaneously, the testing of the Data Acquisition (DAQ) System took place to verify its functionality. Once each component underwent examination and was deemed ready for deployment, the entire proANUBIS setup was transported to CERN LHC Point 1.
\begin{figure}[H]
    \centering
    \begin{subfigure}[b]{0.9\textwidth}
        \includegraphics[width=6.0cm, height=5cm]{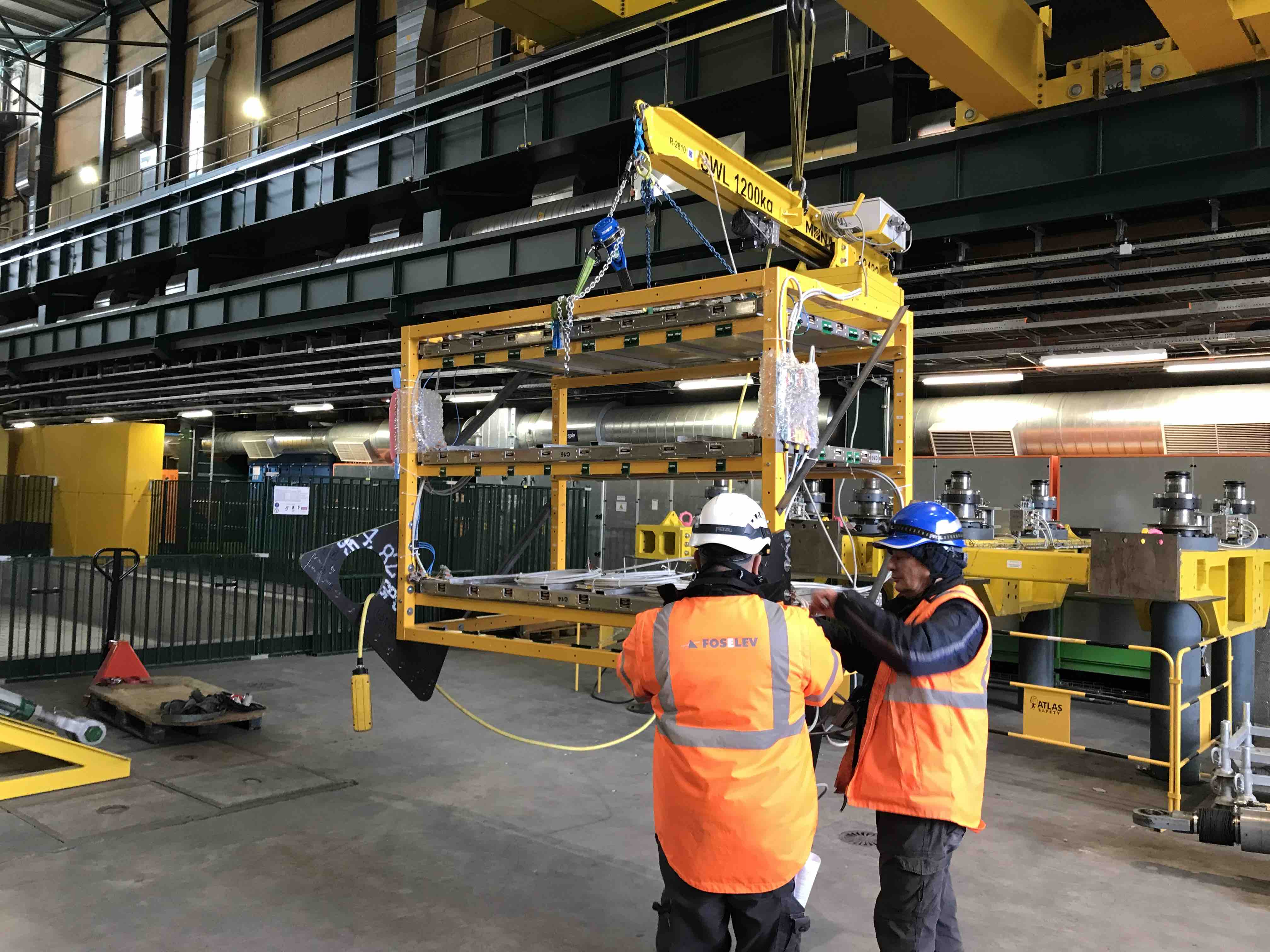} 
        \includegraphics[width=7.0cm, height=5cm]{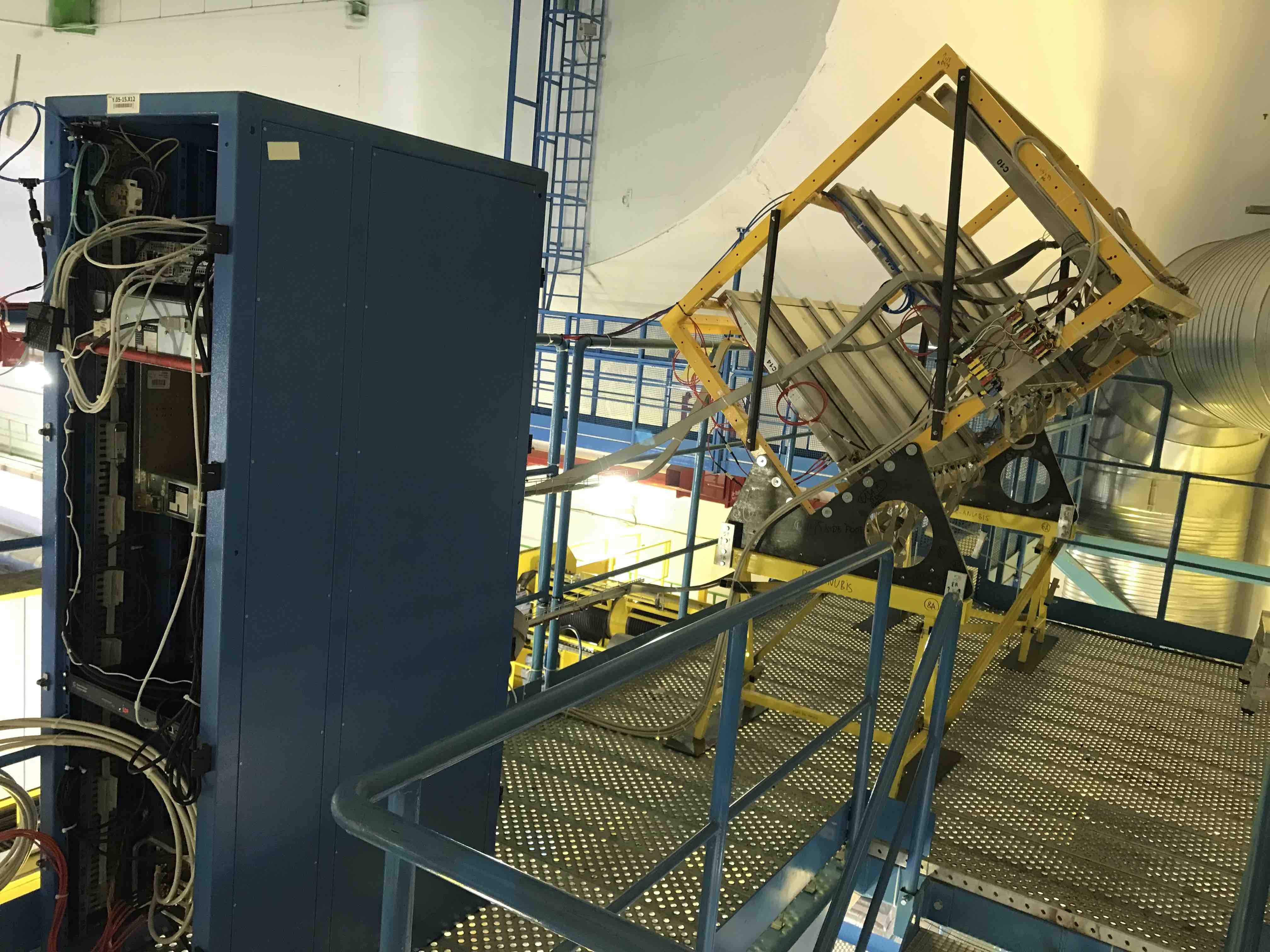} 
    \end{subfigure}
    \caption{(left) A picture captures the moment as proANUBIS is lifted by a crane with the help of ATLAS technical coordination on the surface of SX15 at point-1 of CERN and being lowered through the ATLAS PX14 access shaft.  (right) Picture showcases the installation of proANUBIS at its intended position level-12 side-A of ATLAS underground cavern, as seen along with the Data Acquisition (DAQ) rack.} \label{fig:proANUBIS_installation}
\end{figure}
\vspace{-5pt}

The designated installation site for proANUBIS is located on level 12 side-A of the ATLAS cavern, approximately 80 meters below the earth's surface. With the assistance of ATLAS technical coordination, the proANUBIS detector, along with the DAQ rack, was lowered into position through the PX14 access shafts, using a crane for precise maneuvering. The efforts led to the successful installation of the proANUBIS system, positioned alongside the DAQ rack within the cavern, as illustrated in Figure~\ref{fig:proANUBIS_installation}. These tasks were executed during the Year End Technical Stop (YETS) of 2022, ensuring minimal or no disruption to the operation of the current LHC experiments.
\vspace{-5pt}
\subsection{Preliminary detector commissioning results}
\begin{figure}[H]
    \centering
    \begin{subfigure}[b]{0.5\textwidth}
        \includegraphics[width=7.5cm, height=6.5cm]{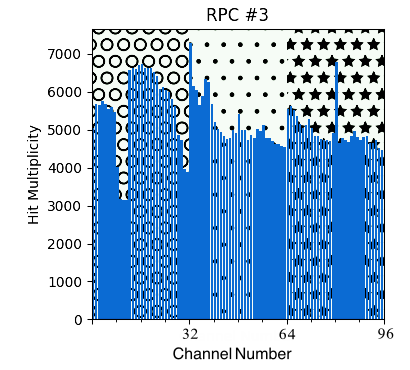} 
    \end{subfigure}
    \caption{An example plot representing the hit multiplicity versus channel number. The blue colour entries represent the number of muon hits on the third RPC detector versus its strip number (or channel number) as recorded by proANUBIS setup. The plot has three sections; circles that represent each of the 32 channels of the eta region while as dots and stars correspond to phi-1 and phi-2 regions of the detector.  The plot corresponds to detector with one eta region and two phi regions being considered as two-dimensional readout  (eta as X strips and phi’s as Y strips) system and this information from all the tracking detectors can be used eventually in particle tracking and background study.} \label{fig:proANUBIS_data_firstLook}
\end{figure}

\vspace{-5pt}
Initial testing of the DAQ system using cosmic rays took place at the CERN BB5 laboratory. Despite limitations imposed by LHC deadlines, this testing yielded valuable insights into the functionality of the DAQ chain.
Figure~\ref{fig:proANUBIS_data_firstLook} provides an initial glimpse of the commissioning data following the installation of the proANUBIS structure in the ATLAS underground cavern. The figure illustrates hit multiplicity versus strip number recorded by the proANUBIS setup for cosmic muons and muons originating from LHC collisions (possibility of having inherent detector and electronics noise in this plot can not be ruled out). Distinction between simultaneous muon events from cosmic rays and LHC collisions is achievable through time-of-flight measurements in offline data analysis.

\vspace{-5pt}
\vspace{-5pt}
\section{Summary and outlook}
\vspace{-5pt}
The successful commissioning and installation of the proANUBIS demonstrator in the ATLAS experiment's underground cavern is an important step, setting a stage for data taking during Run 3 of the LHC. With the prototype now operational, the proANUBIS experiment is poised to undertake studies aimed at providing insights into the expected backgrounds, optimizing detector performance, readout system, and fostering collaboration among various institutions. These efforts are needed to enhance our understanding of the project structure and contribute to the realization of the full-scale ANUBIS detector.


\end{document}